\documentstyle[12pt]{article}

\def\be{\begin{equation}}
\def\ee{\end{equation}}  
\def\bea{\begin{eqnarray}}
\def\eea{\end{eqnarray}}
\def\vk{{\bf k}}
\def\vko{{\bf k}_{1}}
\def\vkt{{\bf k}_{2}}

\begin{document}
\begin{titlepage}
\begin{centering}

\vfill

{\bf Soft photon emission as a sign of sharp transition
in quark-gluon plasma}\footnote {Talk presented at E.S.Fradkin memorial
conference, Moscow, July 2000.\\ This work was supported by Russian
Fund for Fundamental Research,\\ grant 00-02-16101a}

\vspace{1cm}
I.V.Andreev\\
\vspace{0.5cm}
{\it
 Lebedev Physical Institute, 117924, Moscow, Russia}\\

\vspace{3cm}
\end{centering}
\begin{centerline}
{\bf Abstract}
\end{centerline}

\vspace{0.3cm}
Photon emission arising in the course of transition between the states of 
quark-gluon and hadron plasma has been considered. Single-photon 
distributions
and two-photon correlations in central rapidity region have been calculated
for heavy ion collisions at high energies. It has been found that opposite 
side two-photon correlations can serve as a sign of sharp transitions between
the states of strongly interaction matter.
                                          
\vfill \vfill

\end{titlepage}

\section{Introduction}
Forty years ago E.S.Fradkin and his students had calculated the photon
polarization operator in relativistic plasma at finite temperatures~\cite{F}. 
These results will be used here for estimation of a new specific mechanism
of photon production which may appear effective for identification of
transitions between the states of quark and hadron matter in heavy ion
collisions.

The phenomenon under consideration is the photon production in the course 
of
evolution of strongly interacting matter. Let us consider photons existing
in the medium at initial moment $t_0$ having momentum $\bf k$ and energy
$\omega_{in}$. Let the properties of the medium (its dielectric penetrability)
change within time interval $\delta\tau$ so that the final photon energy is
$\omega_f$. As a result of the energy change the production of extra photons
with momenta $\pm\bf k$ takes place these photons having specific
two-photon correlations. Analogous processes were considered for
mesons~\cite{AW,AC,A,ACG} and applied to pion production in high-energy 
heavy
ion collisions~\cite{A1}. 

The conditions for a strong effect are the following: first, the ratio of
the energies ${\omega_{in}/{\omega_f}}$ must not be too close to unity and
second, the transition should be fast enough.

\section{Basic formulation}
Time evolution of the transverse photon creation and annihilation operators
is given by canonical Bogoliubov transformation~\cite{B} which represents
solution of the Hamilton equations and contains two modes with momenta
$\pm\vk$ :
\bea
a(\vk,t)=u(\vk)a(\vk,0)+v(\vk)a^{\dag}(-\vk,0), \nonumber \\
a^{\dag}(-\vk,t)=v(\vk)a(\vk,0)+u(\vk)a^{\dag}(-\vk,0),
\label{eq:1}
\eea
(polarizations are omitted for a moment). Here Bogoliubov coefficients $u,v$
satisfy equation
\be
|u(\vk)|^{2}-|v(\vk)|^{2}=1
\label{eq:2}
\ee
preserving canonical commutation relations and the limit $t\rightarrow\infty$
must be taken. Physically the process under consideration is analogous to
parametric excitation of quantum oscillators. It was considered in more
details earlier~\cite{A}. The Bogoliubov coefficients
$u(\vk)$, $v(\vk)$ are taken to be real valued and $k=|\vk|$ dependent. So we
use a parametrization
\be
u(\vk)=\cosh r(k), \qquad v(\vk)=\sinh r(k)
\label{eq:3}
\ee
thus introducing evolution parameter $r(k)$.

To get feeling of the main features of the evolution effect (and for further
references and comparison) let us formulate a simple model -- fast 
simultaneous
break-up of large homogeneous system at rest~\cite{A1}. In this case the
resulting single-particle momentum distribution can be written in a simple 
form
\bea
\frac{dN}{d^{3}k}=\langle a^{\dag}(\vk)a(\vk)\rangle|_{t\rightarrow\infty}
=\frac{V}{(2\pi)^3}\left[ n(k)\cosh(2r(k))+\sinh^{2}r(k)\right]
\label{eq:4}
\eea
(for single polarization) where $V$ is the volume of the system and $n(k)$
is the level occupation number at $t=0$. The first term in {\it rhs} of Eq.(4)
describes amplification of existed particles and the second term describes
the contribution arising due to rearrangement of the ground state of the
system in the course of the transition.

The transition effect is better seen in particle correlations.Two-particle
inclusive cross-section is given here by
\bea
\frac{1}{\sigma}\frac{d^{2}\sigma}{d^{3}k_{1}d^{3}k_{2}}
=\langle a^{\dag}_{1}a^{\dag}_{2}a_{1}a_{2}\rangle
=\langle a^{\dag}_{1}a_{1}\rangle\langle a^{\dag}_{2}a_{2}\rangle
+\langle a^{\dag}_{1}a_{2}\rangle\langle a^{\dag}_{2}a_{1}\rangle
+\langle a^{\dag}_{1}a^{\dag}_{2}\rangle\langle a_{1}a_{2}\rangle
\label{eq:5}
\eea
The fist term in {\it rhs} of Eq.(5) is the product of single-particle distributions,
the second term gives the usual Hanbury Brown-Twiss effect (HBT) and the 
third
term is essential if time evolution takes place giving opposite side photon
correlations (see below).
The correlators in Eq.(5) in the case under consideration have the form:
\be
\langle a^{\dag}(\vko)a(\vkt)\rangle
=\left[ n(k)+(2n(k)+1)\sinh^{2}r(k)\right]\frac{V}{(2\pi)^{3}}G(\vko-\vkt),
\label{eq:6}
\ee
\be
\langle a(\vko)b(\vkt)\rangle
=\sinh2r(k)\left[n(k)+\frac12\right]\frac{V}{(2\pi)^{3}} G(\vko+\vkt)
\label{eq:7}
\ee
where $G(\vko\pm\vkt)$ represents normalized Fourier transform of the source
volume at break-up stage $(G(0)=1)$.
It is sharply peaked function of $\vko\pm\vkt$ (at zero momentum) having
characteristic scale of the order of inverse size of the source, this scale
being much less than characteristic scales of photon momentum distribution
$n(k)$ and evolution parameter $r(k)$. So the last two functions may be
evaluated at any of momenta $\vko,\vkt\approx\pm\vk$ (we suggest that
the process is $\vk\to -\vk$ symmetric).

Relative correlation function which is measured in experiment is now given by
\be
C(\vko,\vkt)=1+G^{2}(\vko-\vkt)+R^{2}(k)G^{2}(\vko+\vkt)
\label{eq:8}
\ee
with
\be
R(k)=\frac{\sinh r(k)\cosh r(k)(2n(k)+1)}
                  {\sinh^{2}r(k)(2n(k)+1)+n(k)}
\label{eq:9}
\ee
As it can be seen from Eqs.(8-9), HBT effect is given simply by the form-
factor
$G(\vko-\vkt)$ in this model whereas the transition effect depends strongly
on evolution parameter $r(k)$. In turn $r(k)$ depends on time duration 
$\delta\tau$ of the transition. For very small characteristic times
$\delta\tau$ the expression for $r(k)$ is universal~\cite{AW},
\be  
r(k)=\frac{1}{2}\ln\left(\frac{\omega_{f}(k)}{\omega_{in}(k)}\right),\qquad
 \omega\delta\tau\ll 1   
\label{eq:10}
\ee
where $\omega_{in}$ and $\omega_f$ are particle energies before and after 
the
transition. For larger $\delta\tau$ the evolution parameter lessens. 
In general we expect that it falls exponentially at large $\omega\delta\tau $
if the time dependence of the energy in the course of transition has no
singularities at real times. So for large $\omega\delta\tau$  we shall use an 
exponentially falling expression 
motivated by solvable model expression~\cite{A}. Below, after necessary
modification, we apply the above consideration to photon production in heavy
ion collisions.

\section{Photons in plasma}
Spectrum of photons in plasma is given by dispersion equation
\be
\omega^{2}_{k}=k^{2}+\Pi(\omega_{k},k,T,\mu,m)
\label{eq:11}
\ee
Here $\Pi$ is the polarization operator for transverse photons dependent on
temperature $T$, chemical potential $\mu$ and the mass $m$ of charged 
particles. Below we use an approximate form extracted from original
expression~\cite{F}:
\be
\Pi=\omega^{2}_{a}\left[ 1-\frac{\omega^{2}-k^{2}}{k^{2}}\left( \ln \left(
 \frac{\omega+vk}{\omega-vk} \right)-1 \right) \right]
\label{eq:12}
\ee
with
\be
\omega^{2}_{a}=\frac{4g\alpha T^{2}}{\pi}\int_{m/T}^{\infty}dx
\left(x^{2}-\frac{m^{2}}{T^{2}}\right)^{1/2}n_{F}(x,\mu/T)
\label{eq:13}
\ee
where $\alpha=1/137$, $v^2$ is the averaged velocity squared of the charged
particles in the plasma, factor $g$ takes into account the number of the
particle kinds and their electric charges ($g=5/3$ for $u,d$ quarks) and
$n_{F}$ is the occupation number of the charged particles (Fermi 
distribution).
Polarization operator for scalar charged particles is approximately a half
of that for fermions with substitution of Bose distribution for Fermi
distribution. Evidently the polarization operator plays the role of (momentum
dependent) photon termal mass squared $m^{2}_{\gamma}$.

We calculated the polarization operator and photon spectrum for three 
possible
kinds of plasma: quark-gluon plasma (QGP) with $u,d$ light quarks, 
constituent
quark ($m=350 MeV$)-pion plasma and hadronic (pions and nucleons) 
plasma.
Chemical potential (baryonic one) was taken to be equal to $100 MeV$ per 
quark
corresponding to typical value for SPS energies. The temperature was taken
to be equal to $140 MeV$ (see below). 

The evolution parameter $r(k)$ for photons is determined through photon
energy $\omega(k)$. For small momenta $k$ the parameter $r(k)$ is well
approximated by simple expression (to be used for $k_{T}<40 MeV$)
\be
r(k)=\frac{m^{2}_{\gamma1}-m^{2}_{\gamma2}}{4(\langle 
m^{2}_{\gamma}\rangle
+k^{2})}=\frac{\delta m^{2}_{\gamma}}{4\langle\omega^{2}_{k}\rangle} ,
\qquad k\delta\tau\ll1
\label{eq:14}
\ee
where $m^{2}_{\gamma i}$ are photon termal masses squared at both sides 
of
the transition and $\langle m^{2}_{\gamma}\rangle$ is their average mass 
squared 
,cf Eq.(10). At $k=0$ the values of $\delta m^2_\gamma$ are equal to 289, 
178
and 106 (in $MeV^2$ units) for QGP-hadron, QGP-valon and valon-hadron
transitions correspondingly. Corresponding values of zero momentum 
evolution
parameter $r(0)$ are 0.330, 0.154 and 0.178.

Higher momentum behaviour of $r(k)$ (to be used for $k_{T}>40 MeV$) is 
taken
in the form
\be
r(k)=\frac{\delta m^{2}_{\gamma}}{4k^2}\exp \left(-\frac{\pi}{2}k\delta\tau\right)
\label{eq:15}
\ee
where $\delta\tau$ is time duration of the transition. The Eq.(15) is a simple
version of the expression given by the solvable model~\cite{A} which is
sewed together with Eq.(14) giving a monotonically decreasing function of the
momentum. Below Eqs.(14-15) will be used for estimation of the transition 
effect
in heavy ion collisions. Only QGP-hadron transition will be calculated.
In view of fact that evolution parameter $r(k)$ appeared to be small number  
at all momenta $k$, all expressios will be taken in the lowest order in $r(k)$.

\section{Transition effect in heavy ion collisions}
Let us now apply the above consideratins to photon production in heavy ion
collisions. Let us suggest that the quark-gluon plasma is formed at the
initial stage of the ion collision. Let the plasma undergoes expansion and
cooling. The expansion is taken to be longitudinal and boost 
invariant~\cite{BJ}.
Recent lattice calculations~\cite{LAT} show rather low critical temperature
of the deconfinement and chiral phase transition, $T_c\approx 150 MeV$
as well as sharp drop of the pressure when the temperature approachs $T_c$
thus provocing instability in the presence of overcooling. So we do not
expect long-living mixed phase and consider fast transition from quark to
hadron matter with characteristic transition proper time duration $\delta\tau$
of the order of $1 fm/c$.

To calculate the transition effect one must shift to rest frame of each moving
element of the system and integrate over proper times and space-time 
rapidities
of the elements. Then single-photon distribution in central rapidity region
$y=0$ reads:

\begin{eqnarray}
\frac{dN}{d^{2}k_{T}dy}\Bigl|_{y=0}\Bigr.=I_{QGP}+I^{(1)}_{tr}\nonumber\\
=\int\tau d\tau\int d\eta\int d^{2}x_{T}(p_{0}\frac{dR_{\gamma}}{d^{3}p})
+\int d\eta\int d^{2}x_{T}\frac{2p\tau_{c}}{(2\pi)^{3}}r^{2}(p)
\label{eq:16}
\end{eqnarray}                       
with $p=k_{T}\cosh\eta$.

The first term in rhs of Eq.(16) describes photon production from hot 
quark-gluon plasma. Here $R_{\gamma}$ is the QGP production rate per unit
four-volume in the rest frame of the matter~\cite{KLS}:

\begin{equation}
p_{0}\frac{dR_{\gamma}}{d^{3}p}=\frac{5\alpha\alpha_{s}}{18\pi^{2}}T^{2}
\exp(-p/T)\ln(1+\frac{\kappa p}{T})
\label{eq:17}
\end{equation}

with $\alpha=1/137$, $\alpha_{s}=0.4$, $\kappa=0,58$. It can be used  also 
for
hadron gas as its uncertainity is larger than the difference between the
first-order QGP and hadron gas production rates~\cite{NKL}. Contribution 
from 
hadronic resonances are not considered here. The second term in {\it rhs} of 
Eq.(16)
describes photon production due to transition from QGP to hadrons in the
vicinity of proper time $\tau_{c}$, cf Eq.(4). The time duration of the
transition is taken to be small in this term in comparison with total time
duration of photon production process.

The photon production rate in Eq.(16) can be expressed through photon 
occupation number $n(k)$:
\begin{equation}
p_{0}\frac{dR_{\gamma}}{d^{3}p}=\frac{2k_{T}}{(2\pi)^3}\frac{dn(k)}{d\tau}
\label{eq:18}
\end{equation}
(with two polarizations included). Taking Eq.(18) into account one can see
that if the velocities of the volume elements, as well as proper time interval
in the first term in {\it rhs} of Eq.(16) are small then Eq.(16) is reduced to
Eq.(4) as it should be. The photon occupation number $n(k)$ in Eq.(18)
appears to be small numerically,
$$
n(k)\ll 1
$$
That means in particular that transition radiation is dominated not by the
photon amplification but by the ground state rearrangement (cf Eqs.(4-8)).

As the last step one must specify temperature evolution. We suggest that the
temperature depends on proper time of the volume element with power-like
dependence:
\begin{equation}
(T/T_{0})=(\tau/\tau_{0})^{-1/b}
\label{eq:19}
\end{equation}
where $\tau_{0}$ and $T_{0}$ are initial proper time and initial temperature.
For final estimation we use $b=3$ typical for hydrodynamical picture and
choose low transition temperature $T_{c}=140 MeV$. After transition the 
photons
live some time in hadronic medium and we suggest termal momentum 
distribution
of the hadrons (modified by the expansion of the system). We neglect termal
photon production below $T_{c}$ and do not introduce a special freese-out
temperature.

Below we will be interested in rather low photon transverse momenta $k_{T}$
(up to $500 MeV$) where transition effect is more pronounced. In this
momentum region the QGP production term $I_{QGP}$ depends mainly on 
final
temperature $T_c$. For two main variants of initial conditions used in
literature~\cite{NFS}: $\tau_{0}T_{0}=1, T_{0}/T_{c}=3/2{}, \tau_{0}=1 fm/c $
and $\tau_{0}T_{0}=1/3, T_{0}/T_{c}=5/2{}, \tau_{0}=0.2 fm/c$ a variable factor
in $I_{QGP}$ changes
inessentially (from 5.06 to 4.37 for $b=3$) and we will use for this factor
an average value 4.7 in our estimations. The transition proper time $\tau_c$
also changes inessentially for these two variants of initial conditions 
being $3.00 fm/c$ and $3.02 fm/c$ correspondingly.

The transition contribution $I^{(1)}_{tr}$ in Eq.(16) appears essential
only at very small momenta $k_T$. So dealing with single-photon 
distributions
one can use Eq.(14) for evolution parameter $r(k)$. The resulting relative
strength of transition radiation
\begin{equation}
R_{1}(k_{T})=I^{(1)}_{tr}/I_{QGP}
\label{eq:20}
\end{equation}
appears sizable only in the momentum region $k_{T}\leq (15-20) Mev$ 
independently of the time duration of the transition.
 $R_{1}$ reachs 4.44 at $k_{T}=0$ and falls down to 0.06 at $k_{T}=40 MeV$.
Because of high background effects in this momentum region the single-
photon
transition effect should be difficult to observe experimentally.
 

Mach better the transition effect is seen in photon correlations (cf Eqs.(5-9))
where it is first order effect with respect to $r(k)$. Let us note that HBT
effect for photons has now more complicated form than that in Eq.(8) because
of finite time duration~\cite{APW} of the process of photon emission from 
QGP  
and it will not be exposed here. We consider only the transition effect (the
third term in Eqs.(5,8) which gives opposite side correlations) estimating its 
contribution to two-photon correlation function in central rapidity region.
Suggesting fast transition we can evaluate the contribution in the vicinity
of fixed proper time $\tau_c$. So one only has to shift to the rest frame
of each element of the expanding volume and perform $\eta$-integration. Then
the extention of the correlator in Eq.(7) to the case of expanding volume
takes the form:
\begin{equation}
2k\langle a({\bf k}_{1})a({\bf k}_{2})\rangle=G({\bf k}_{1T}+{\bf k}_{2T})
I^{(2)}_{tr}
\label{eq:21}
\end{equation}
with
\begin{equation}
I^{(2)}_{tr}=\int d^{2}x_{T}\int d\eta \frac{2\tau_{c}k_{T}\cosh\eta}{(2\pi)^3}
r(k_{T}\cosh\eta)
\label{eq:22}
\end{equation}
where we neglected $n(k)$ in comparison with unity (see above). Therefore
the normalized two-photon correlation function is given by (cf Eqs.(8-9))
\begin{equation}
C({\bf k}_{1T},{\bf k}_{2T})|_{y_{1}=y_{2}=0} =1+C_{HBT}
+R^{2}(k_{T})G^{2}({\bf k}_{1T}+{\bf k}_{2T})
\label{eq:23}
\end{equation}
with
\begin{equation}
R(k_{T})=\frac{I^{(2)}_{tr}}{I_{QGP}+I^{(1)}_{tr}}
\label{eq:24}
\end{equation}

We calculated the ratio $R(k_{T})$ for different transition times $\delta\tau
=0 fm/c,\\ 0.5 fm/c, 1 fm/c$ up to $k_{T}=500 MeV$.
%
In the region $k_{T}<100 MeV$ the ratio $R$ is sizable for all these
$\delta\tau$ 
being equal 4.94 at $k_{T}=0$, reaching maximal value $R\sim 6$ at 
$k_{T}\sim
20 MeV$ and falling down at $k_{T}=100 MeV$
to $R=1.78$ for $\delta\tau=0$, $R=0.95$ for
$\delta\tau=0.5 fm/c$, $R=0.55$ for $\delta\tau=1 fm/c$ .
At larger transverse momenta the behaviour of the ratio $R$ depends strongly
on transition time $\delta\tau$: in $k_{T}$ interval $(200-500) MeV$
the ratio $R(k_{T})$ rises from 1.3 to 3.9 for $\delta\tau=0$, it is
approximately constant ($R$=0.3) for $\delta\tau=0.5 fm/c$ and it decreases
from 0.15 to 0.03 for $\delta\tau=1 fm/c$.
On the whole it seems that the measurement of $R(k_{T})$ gives a possibility
to identify the transition effect, especially if $\delta\tau\leq 0.5 fm/c $

\section{Conclusion}
Estimation of photon emission accompaning transition between quark-gluon 
and
hadron states of matter in heavy ion collisions shows that opposite side
photon correlations can serve as a sign of the transition if transition
time is small enough.

\end{document}